\documentclass[11pt]{article} 
\usepackage{rldmsubmit,palatino}
\usepackage{graphicx}

\usepackage{url}

\usepackage{amsmath}
\usepackage{amssymb}
\usepackage{mathletters}

\usepackage{algorithm}
\usepackage{algorithmic}

\usepackage{subcaption}

\title{Query Completion Using Bandits for Engines Aggregation}

\author{
Audrey Durand \\
Universit\'e Laval \\
\texttt{audrey.durand.2@ulaval.ca} \\
\And
Jean-Alexandre Beaumont \\
Universit\'e Laval \\
\texttt{jean-alexandre.beaumont.1@ulaval.ca} \\
\AND
Christian Gagn\'e \\
Universit\'e Laval \\
\texttt{christian.gagne@gel.ulaval.ca} \\
\And
Michel Lemay \\
Coveo \\
\texttt{mlemay@coveo.com} \\
\And
S\'ebastien Paquet \\
Coveo \\
\texttt{spaquet@coveo.com}
}

\begin{document}

\maketitle

\begin{abstract}
Assisting users by suggesting completed queries as they type is a common feature of search systems known as query auto-completion. A query auto-completion engine may use prior signals and available information (e.g., user is anonymous, user has a history, user visited the site before the search or not, etc.) in order to improve its recommendations. There are many possible strategies for query auto-completion and a challenge is to design one optimal engine that considers and uses all available information. When different strategies are used to produce the suggestions, it becomes hard to rank these heterogeneous suggestions. An alternative strategy could be to aggregate several engines in order to enhance the diversity of recommendations by combining the capacity of each engine to digest available information differently, while keeping the simplicity of each engine. The main objective of this research is therefore to find such mixture of query completion engines that would beat any engine taken alone. We tackle this problem under the bandits setting and evaluate four strategies to overcome this challenge. Experiments conducted on three real datasets show that a mixture of engines can outperform a single engine.
\end{abstract}

\keywords{
query; completion; aggregation; bandits
}

\acknowledgements{Thanks to Pascal Soucy and Steffen Kirres from Coveo who helped implementing the query completion engines used in this project. Thanks to Coveo for providing the data and the infrastructure and to the Natural Sciences and Engineering Research Council of Canada (NSERC) for the research grant EGP 492531-15.}

\startmain 

\section{Introduction}

A common feature in search systems is to assist users in formulating their queries by suggesting completed queries as they type. This is known as query auto-completion (QAC). The typical QAC problem consists in providing a user with the top-$K$ completion suggestions taken from a set of possible suggestions, given a user-provided query prefix and using prior signals for ranking completion suggestions~\cite{Whiting2014}. For example, a QAC engine could generate for the user input ``que'' the suggestions 1) ``query'', 2) ``question'', and 3) ``query results''.
It is an important feature that provides many advantages: users can write queries faster, write more precise and complete queries, use the right vocabulary, avoid typos, and execute queries that have proven to be successful in the past. Moreover, it has the side effect of standardizing the queries, which helps an adaptive search system learning the best documents to return for each query.

Much work has been done in order to design good query completion engines that consider contextual information (e.g.~\cite{Bar-Yossef2011,Duan2011,Shokouhi2013,Strizhevskaya2012,Whiting2014}), which might include the status of the user (anonymous or logged in), user history, Web pages visited prior to the search, and much more. There are many possible strategies for QAC and a challenge is to design an engine that uses all available information in order to recommend diverse relevant suggestions given all this knowledge. Inspired by resource aggregation techniques~\cite{Selberg1997}, a strategy could be to aggregate several engines instead of aiming for one optimal engine. More specifically, each position of the suggestions list could be assigned to an engine and filled with a suggestion provided by this engine. This could enhance the diversity of suggestions by combining the strengths of different engines, each using the contextual information in its own specific way. The main objective of this research is thus to find a mixture of QAC engines that would beat any engine taken alone trying to consider all information at once. Constraint is that the learning process must be performed online, that is without an a priori learning phase before deployment.
To achieve these objectives, we propose bandits-based techniques adapted from previous work.

\paragraph{Related Works}

Bandits-based techniques have previously been considered to tackle the query suggestion problem, where the goal is to suggest additional queries to a user given its past queries~\cite{Hsieh2015}. Bandits algorithms in this setting were used to learn a mapping from each query to the top-$K$ most relevant other queries. This would lead to a very large model, that is one mapping per possible query, and it would not necessarily be useful since many queries might occur a single time in history. Therefore, it was limited to the most frequent queries, which made sense for the query suggestion problem where full query terms are considered. However, it was found to be limiting in the QAC problem, where the most frequent queries are short sub-queries that are very common among multiple query terms.

Bandits-based techniques have also been considered for the recommendation problem, where the goal is to answer the search query of a user with a list of several items.  Bandits algorithms in this setting were used to learn a mapping from each query to the top-$K$ most relevant links or documents. Previous research mainly addressed the issue of redistributing feedback given the position of click occurrence(s).
The same kind of questions raise for the QAC problem and the models in the following proposed approach are based on techniques from this field.


\section{QAC as Mixture of Engines}

We tackle the QAC problem using a mixture of engines (QAC-ME), which we formalize as follows. Let $\cE$ denote a set of QAC engines. On each time $t$, a list of $M$ auto-completion suggestions is displayed to the user according to the current user-provided query prefix $p_t$. Let a \emph{good suggestion} denote a suggestion that would please the user. The user satisfaction toward a suggestion can be measured through user clicks. Let $c_t \in \{ 1, \dots, M+1 \}$ denote the position of the suggestion that is clicked by the user if any, otherwise $c_t = M+1$. The goal is to maximize the number of user clicks over time.

Let $\cS_{e, t}$ denote the set of suggestions provided by engine $e$ at time $t$ using $p_t$ and possibly other contextual information. Items in $\cS_{e, t}$ are ordered by relevance such that the top-$K$ items correspond to the first $K$ items in the set. We want to assign an engine to each position of the QAC list such that this engine is in charge of providing the suggestion displayed in this position. Let $e_{m, t} \in \cE$ denote the engine designated to fill position $m$ and let $q_{m, t} \in \cS_{e_{m, t}, t}$ denote the suggestion assigned to position $m$. Duplicate suggestions are forbidden, meaning that $q_{m, t}$ is the most relevant suggestion from $e_{m, t}$ such that $q_{m, t} \neq q_{i, t}$ for $i \leq m-1$.  The goal is to design an algorithm that selects the engine to use at each position in order to maximize the probability that the user clicks on any suggestion from the list, that is the probability that $c_t \neq M+1$. Note that it has been observed that the probability of getting a click on an item decays with the rank of the item in a list~\cite{Craswell2008} -- for example, a good suggestion in position~1 as a higher click probability than the same suggestion in position~3.

\section{Approaches}
\label{sec:approach}



%
The ranked model (Alg.~\ref{alg:ranked}) based on the ranked bandits algorithm~\cite{Radlinski2008} for query recommendation handles each position as an independent bandits problem, instantiating one bandits algorithm $\phi_m$ for each position $m$.
The ranked model does not share information from feedback gathered on the same engine placed at different positions.
%
\begin{figure}
    \begin{minipage}[t]{3.6in}
        \begin{algorithm}[H]
            \begin{algorithmic}[1]
                \STATE initialize $\phi_1(\cE), \dots, \phi_M(\cE)$
                \FORALL{episodes $t$}
                    \STATE receive prefix $p_t$ from user
                    \FOR{$m = 1, \dots, M$}
                        \STATE $e_{m, t} \leftarrow \text{select}(\phi_m)$
                        \REPEAT
                            \STATE $q_{m, t} \leftarrow$ suggestion of engine $e_{m, t}$ for prefix $p_t$
                        \UNTIL{$q_{m, t} \not\in \{ q_{1, t}, \dots, q_{m-1, t} \}$}
                    \ENDFOR
                    \STATE display $\{ q_{1, t}, \dots, q_{m, t} \}$ to user and get click index $c_t$
                    \STATE update $\phi_{c_t}$ with outcome 1 for action $e_{c_t, t}$
                    \STATE update $\phi_m$ with outcome 0 for action $e_{m, t},~\forall m \neq c_t$
                \ENDFOR
            \end{algorithmic}
            \caption{Ranked Bandits for QAC-ME}
        \label{alg:ranked}
        \end{algorithm}
    \end{minipage}
    \hfill
    \begin{minipage}[t]{3.6in}
        \begin{algorithm}[H]
            \begin{algorithmic}[1]
                \STATE initialize $\phi(\cE)$
                \FORALL{episodes $t$}
                    \STATE receive prefix $p_t$ from user
                    \FOR{$m = 1, \dots, M$}
                        \STATE $e_{m, t} \leftarrow \text{select}(\phi)$
                        \REPEAT
                            \STATE $q_{m, t} \leftarrow$ suggestion of engine $e_{m, t}$ for prefix $p_t$
                        \UNTIL{$q_{m, t} \not\in \{ q_{1, t}, \dots, q_{m-1, t} \}$}
                    \ENDFOR
                    \STATE display $\{ q_{1, t}, \dots, q_{m, t} \}$ to user and get click index $c_t$
                    \STATE update $\phi$ with outcome $1$ for action $e_{c_t, t}$
                    \STATE update $\phi$ with outcome $0$ for action $e_{m, t},~\forall m < c_t$
                \ENDFOR
            \end{algorithmic}
            \caption{Cascade Bandits for QAC-ME}
        \label{alg:cascade}
        \end{algorithm}
    \end{minipage}
\end{figure}
In contrast, the cascade model (Alg.~\ref{alg:cascade}) based on the cascade bandits algorithm~\cite{Kveton2015} for query recommendation uses one single bandits algorithm $\phi$ for the whole setting.
The cascade model merges all information obtained for a given engine regardless of the location of the engine when feedback was gathered. This should speed up the learning process but this also assumes independence between engines performance and their location in the list, which might not be true in practice.
Note that neither $\phi_m$ (ranked) nor $\phi$ (cascade) considers which engines are assigned to positions 1 to $m-1$, or which suggestions are placed in these positions when selecting $e_{m,t}$.


Let the \emph{rank} of the suggestion for engine $e$ denote its index in $\cS_{e, t}$. Obviously, if engine $e$ is asked to fill position $m$ (Algs.~\ref{alg:ranked} and~\ref{alg:cascade}, line 7), it will recommend its most relevant suggestion, that is the first suggestion in $\cS_{e, t}$, or rank 1. However, because showing duplicate suggestions to the user is forbidden, engine $e$ is asked for its next suggestion until a new, unique, suggestion is provided (Algs.~\ref{alg:ranked} and~\ref{alg:cascade}, line 8). Given that $M$ positions must be filled, engines might be forced to recommend up to their $M$-th best suggestion\footnote{We assume that there are no duplicates among the suggestions $\cS_{e, t}$ of a given engine $e$.}. An easy example is when the same engine is assigned to fill all $M$ positions. It will obviously recommend its top-$M$ suggestions.

It is natural to assume that the probability of showing a good suggestion for an engine may vary given the rank of the suggestion that is actually shown. We address this concern by expliciting the rank of each suggestion placed by a given engine. Let $j_{e, t}(m)$ denote the rank of the most relevant recommendation $q$ from engine $e$ such that $q \neq q_{i, t}$ for $i \leq m-1$. Then $\cA_t(m) = \{ (e, j_{e, t}(m)) \}_{e \in \cE}$ denotes the set of all (engine, rank) tuples \emph{available} for selection at position $m$ on episode $t$. The available actions for filling the first position always corresponds to each engine giving its first rank suggestion: $\cA_t(1) = \{ (e, 1) \}_{e \in \cE}$. The available actions for filling further positions depend on which engines have been used in previous positions and what suggestions they have provided.

Algs.~\ref{alg:ranked_explicit} and~\ref{alg:cascade_explicit} respectively extend Algs.~\ref{alg:ranked} and~\ref{alg:cascade} to the explicit suggestion rank setting. Instead of learning a general outcome distribution per engine, the refined learning process aims at learning one outcome distribution for each suggestion rank per engine. Notice that even though the explicit cascade model still has only one single bandits algorithm that manages all positions, its set of available actions differs from one position to another.
\begin{figure}
    \begin{minipage}[t]{3.6in}
        \begin{algorithm}[H]
            \begin{algorithmic}[1]
                \STATE initialize $\phi_1(\cE), \dots, \phi_M(\cE)$
                \FORALL{episode $t$}
                    \STATE receive prefix $p_t$ from user
                    \FOR{$m = 1, \dots, M$}
                        \STATE $(e_{m, t}, i_{m, t}) \leftarrow \text{select}(\phi_m, \cA_t(m))$
                        \STATE $q_{m, t} \leftarrow$ suggestion $i_{m, t}$ of engine $e_{m, t}$ for prefix $p_t$
                    \ENDFOR
                    \STATE display $\{ q_{1, t}, \dots, q_{m, t} \}$ to user and get click index $c_t$
                    \STATE update $\phi_{c_t}$ with outcome $1$ for action $(e_{c_t, t}, i_{c_t, t})$
                    \STATE update $\phi_m$ with outcome $0$ for action $(e_{m, t}, i_{m, t})$, $\forall m \neq c_t$
                \ENDFOR
            \end{algorithmic}
            \caption{Explicit Ranked Bandits for QAC-ME}
            \label{alg:ranked_explicit}
        \end{algorithm}
    \end{minipage}
    \hfill
    \begin{minipage}[t]{3.6in}
        \begin{algorithm}[H]
            \begin{algorithmic}[1]
                \STATE initialize $\phi(\cE)$
                \FORALL{episode $t$}
                    \STATE receive prefix $p_t$ from user
                    \FOR{$m = 1, \dots, M$}
                        \STATE $(e_{m, t}, i_{m, t}) \leftarrow \text{select}(\phi, \cA_t(m))$
                        \STATE $q_{m, t} \leftarrow$ suggestion $i_{m, t}$ of engine $e_{m, t}$ for prefix $p_t$
                    \ENDFOR
                    \STATE display $\{ q_{1, t}, \dots, q_{m, t} \}$ to user and get click index $c_t$
                    \STATE update $\phi$ with outcome $1$ for action $(e_{c_t, t}, i_{c_t, t})$
                    \STATE update $\phi$ with outcome $0$ for action $(e_{m, t}, i_{m, t})$,\\$\forall m < c_t$
                \ENDFOR
            \end{algorithmic}
            \caption{Explicit Cascade Bandits for QAC-ME}
        \label{alg:cascade_explicit}
        \end{algorithm}
    \end{minipage}
\end{figure}
These explicit variants might converge slower than their original, non-explicit, counterpart because they share less information.
%
However, even though observation gathering takes more time, we would expect these explicit variants to be more robust to suggestions skipped when avoiding duplicates and to be more robust to high performance variance across suggestion ranks in engines.

\section{Application}

We tackle the problem of learning a mixture of four \emph{real} engines for filling $M=5$ positions of a QAC field using three real datasets built by taking full-length queries performed on websites over a one month period and splitting them into (query prefix, full query) tuples. Three different clients were chosen for their diversity and representativeness of real life situations:
\begin{itemize}
    \item Dataset 1: website with few traffic (13k queries per month);
    \item Dataset 2: website with high traffic (1.2M queries per month) and long queries on average;
    \item Dataset 3: website with high traffic (1.1M queries per month) and short queries on average.
\end{itemize}

The Ranked (Alg.~\ref{alg:ranked}), Ranked Explicit (Alg.~\ref{alg:ranked_explicit}), Cascade (Alg.~\ref{alg:cascade}), and Cascade Explicit (Alg.~\ref{alg:cascade_explicit}) strategies are compared against two baselines: the basic engine that is currently deployed by the company and a random mixture assigning engines at random to each position. Note that the basic engine is part of the four engines available for the mixture. Each engine is designed to consider different contextual information such as user history, previous queries (for this user and all users), most popular searches, dictionary entries, and many more.
%
The well-known Thompson sampling (TS)~\cite{Thompson1933} bandits algorithm $\phi$ is used. TS maintains a posterior distribution on the outcome probability of each action given past observations and selects actions according to their probability of being optimal using a sampling procedure. It has been considered previously for the query recommendation problem~\cite{Hsieh2015}. Bernoulli priors and Beta posteriors are used here.
%

Approaches are compared based on the average number of clicks they manage to gather after a trial period and the corresponding increase number of clicks w.r.t. the currently deployed solution, that is the basic engine without mixture. The experiment is run over 10,000 query prefixes (episodes) and each experiment is repeated five times.
On episode $t$, a tuple $(p_t, z_t)$ is sampled from the dataset, were $z_t$ is the full query. We consider that a user click happens in position $m$ if $q_{m, t} = z_t$.
%
Tab.~\ref{tab:experiments:application:results} shows the results (averaged over the five runs) for the three datasets.
\begin{table}[t]
    \centering
    \begin{tabular}{lcrcrcr}
    \hline
     & \multicolumn{2}{c}{Dataset 1} & \multicolumn{2}{c}{Dataset 2} & \multicolumn{2}{c}{Dataset 3} \\
     & Clicks & Increase (\%) & Clicks & Increase (\%) & Clicks & Increase (\%) \\
    \hline
    Current (Basic) & 2104 & -- & 4379 & -- & 6113 & -- \\
    Random & 2905 & 38.07 & 4373 & -0.14 & 5762 & -5.74 \\
    Ranked & 3067 & 45.77 & 4656 & 6.33 & 6094 & -0.31 \\
    Ranked Explicit & 3118 & 48.19 & 4917 & 12.29 & 5900 & -3.48 \\
    Cascade & 2966 & 40.97 & 4457 & 1.78 & 6106 & -0.11 \\
    Cascade Explicit & 3122 & 48.38 & 4924 & 12.45 & 5972 & -2.31 \\
    \hline
    \end{tabular}
\caption{Number of clicks and percentage of increase w.r.t. the basic strategy after 10,000 episodes.}
\label{tab:experiments:application:results}
\end{table}
%
%

We observe that Cascade Explicit and Ranked Explicit manage to gather much more clicks than the other strategies on datasets~1 and~2, leading to large increases with respect to the original basic strategy (up to 48\%). Even the random strategy performs really well compared to the current basic engine on dataset 1. This highlights the potential benefits of a mixture for providing a diverse suggestions list to the user. The improved performance of explicit algorithms compared with their non-explicit variants leads us to believe hat there is a benefit in modeling independently the expected click probability for each rank. We also observe that Ranked does not beat Cascade when the rank is explicit.

\begin{figure}
    \captionsetup{skip=7pt}
    \centering
    \begin{subfigure}[b]{0.49\textwidth}
        \includegraphics{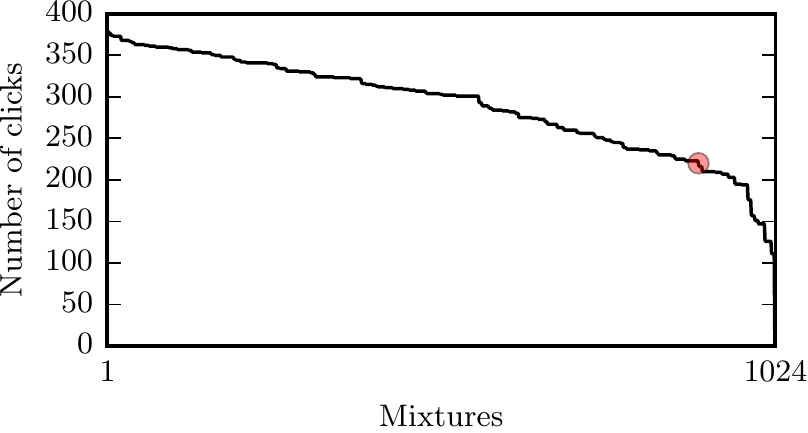}
        \caption{Dataset 1}
    \end{subfigure}
    \begin{subfigure}[b]{0.49\textwidth}
        \includegraphics{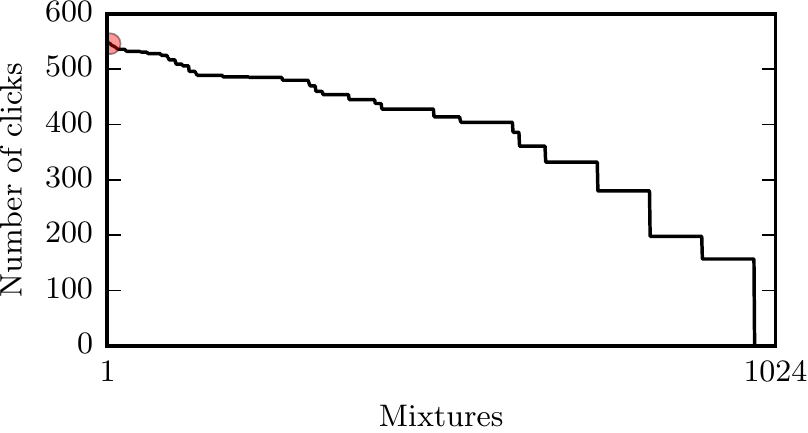}
        \caption{Dataset 3}
    \end{subfigure}
    \caption{Number of clicks per mixture in decreasing order. The red dot indicates the position of the basic strategy.}
\label{fig:experiments:application:allpermutations}
\end{figure}

On dataset~3, it appears that none the strategies are able to beat to the basic strategy. This was expected given that this dataset was generated from data acquired using this engine running and proposing auto-completions to users. In order to validate this hypothesis, we perform additional experiments where we run each possible mixture of the four engines in $M = 5$ positions, that is 1024 mixtures, over 1000 query prefixes (episodes)\footnote{The whole set of 10,000 query prefixes was not used for computing reasons.}. Figure~\ref{fig:experiments:application:allpermutations} shows the total number of clicks obtained with each mixture on datasets~1 and~3, where mixtures have been ordered in decreasing number of clicks. Note that a single run per mixture was performed, meaning that these results are noisy and that the ordering of the mixtures is not absolute. The position of the basic strategy is shown by the red dot.
We observe that the basic strategy is far from being optimal on dataset~1, while it is very close to the top (6-th position) on dataset~3. This confirms why none of the proposed strategies could beat the basic strategy on this dataset. We also note that non-explicit algorithms converge faster than their explicit counterparts.
%
\textbf{Note:} Though Cascade Explicit always seems to beat Ranked Explicit, performing a Welch's t-test revealed that the null hypothesis cannot be rejected in this case. Further replications should be performed for additional conclusions.

\section{Conclusion}

These preliminary results show the potential of mixing query completion engines with bandits-based approaches for improving the quality of the suggestions in the QAC problem and that a mixture adapts better to the large range of usage contexts. Bandit algorithms have shown to be efficient, flexible, and fast to learn which engine to use where and when. Fancier bandits frameworks such as sleeping bandits and structured bandits should also be considered for this problem as they might be more adapted to the dynamics of this application than standard bandits. Results also show a limitation of the offline evaluation setting, that is the dependency upon the approach used for gathering data. This should be a motivation for further, online, experiments.

Future work includes an A/B testing of the strategies on the live system as it would allow us to validate the results presented in this paper. It would also allow us to evaluate the bias introduced by offline evaluation and take into account the real click probability decay pattern. The integration of the bandits algorithms in the company QAC product is already planned to replace the basic engine in order to provide heterogeneous suggestions based on the context. Additional experiments will be also conducted to apply similar approaches to document recommendation. Finally, it would be interesting to provide the analysis in order to obtain theoretical guarantees and regret bounds.

\bibliographystyle{abbrv}
\bibliography{references}

\end{document}